\documentclass[submission,copyright,creativecommons]{eptcs}

\usepackage{amsmath}
\usepackage{amssymb}
\usepackage[T1]{fontenc}
\usepackage{graphicx}
\usepackage{stmaryrd}

\usepackage{listings}
\newcommand{\coqinline}[1]{\textsf{#1}}
\newcommand{\denote}[1]{\left\llbracket #1 \right\rrbracket}
\newcommand{\ifthenelse}[3]{\text{if} \ (#1) \ \text{then} \ #2 \ \text{else} \ #3}

\title{A Coq Library of Sets for Defining Denotational Semantics}

\author{
  Qinxiang Cao
  \institute{Shanghai Jiao Tong University,  Shanghai, China}
  \email{caoqinxiang@gmail.com}
  \and
  Xiwei Wu
  \institute{Shanghai Jiao Tong University,  Shanghai, China}
  \email{yashen@sjtu.edu.cn}
  \and
  Yalun Liang
  \institute{Shanghai Jiao Tong University,  Shanghai, China}
  \email{symb@olic.link}
}

\def\authorrunning{Cao, Wu and Liang}

\def\titlerunning{A Coq Library of Sets for Defining Denotational Semantics}

\begin{document}

\maketitle

\begin{abstract}
Sets and relations are very useful concepts for defining denotational semantics.
In the Coq proof assistant, curried functions to \coqinline{Prop} are used to represent sets and relations, e.g. \coqinline{A -> Prop}, \coqinline{A -> B -> Prop}, \textsf{A -> B -> C -> Prop}, etc.
Further, the membership relation can be encoded by function applications, e.g. 
\coqinline{X a} represents $a \in X$ if \coqinline{X: A -> Prop}.
This is very convenient for developing formal definitions and proofs for professional users, but it makes propositions more difficult to read for non-professional users, e.g. students of a program semantics course.
We develop a small Coq library of sets and relations so that standard math notations can be used when teaching denotational semantics of simple imperative languages.
This library is developed using Coq's type class system. It brings about zero proof-term overhead comparing with the existing formalization of sets.
\end{abstract}

\section{Introduction}
\label{sect:introduction}

Theorem provers are very useful in teaching program semantics and program logics.
Formal definitions can help students understand different concepts more precisely.
Also, students can get immediate feedback from the theorem prover when writing a machine-checkable proof in their homework.

However, it is not easy to redesign existing courses by adding a theorem prover to it.
To clarify, our objective here is to design course material that focuses on programming language theories. Interactive theorem proving is only used to help teaching.
We cannot spent too much time on how to use a theorem prover and how to turn math concepts into formal definitions.
Thus, it is especially critical to ensure that formal definitions in course material are intuitive to read, otherwise students need to pay significant efforts on learning and understanding the formal language used, in addition to learn their counterpart written in standard math symbols.
When we prepared course material about denotational semantics, we found that the existing Coq formalization of sets and relations cannot be used directly in teaching.

\paragraph{Problems in sets' definitions.}
Typically, if a set \coqinline{A} is formalized as a Coq type, then a subset of \coqinline{A} can be formalized as a function of type \coqinline{A -> Prop}, and set operators like union and intersection can be defined as higher-order functions.
For example\footnote{Coq.Sets.Ensemble \cite{CoqSets} defines subsets like this and its formalization of union is equivalent to our definition here.},
\begin{lstlisting}[mathescape=true]
    Definition union1 (X Y: A -> Prop) :=
      fun a => X a \/ Y a.
    Notation "x $\cup$ y" := (union1 x y).
\end{lstlisting}
Moreover, the membership relation can be encoded by function applications, e.g. 
\coqinline{X a} represents $a \in X$ if \coqinline{X: A -> Prop}.

Similarly, relations from \coqinline{A} to \coqinline{B} (i.e. subsets of \coqinline{A} $\times$ \coqinline{B}) can be formalized as curried functions of type \coqinline{A -> B -> Prop} and the union operator can be formalized as:
\begin{lstlisting}
    Definition union2 (X Y: A -> B -> Prop) :=
      fun a b => X a b \/ Y a b.
\end{lstlisting}
For membership, \coqinline{X a b} represents $(a, b) \in X$ if \coqinline{X: A -> B -> Prop}.
Here, using curried functions instead of its uncurried counterparts of type \coqinline{A * B -> Prop} allows us to avoid verbose construction and destruction of pairs in proofs.

Although curried functions are widely used for representing subsets and relations\footnote{For example, Coq.Relations.Relation\_Definitions \cite{Relation} defines relations from \coqinline{A} to \coqinline{A} by functions of type \coqinline{A -> A -> Prop}.}, it brings two problems.

(a) It is nontrivial to unify \coqinline{union1} and \coqinline{union2} above.
We want to unify them because it is inconvenient to have two different Coq definitions for \emph{union}.
Using those two definitions means that, important properties like union's commutativity and associativity should also have two copies. It can be annoying that we need to choose a correct version every time we try to apply them.
Especially in our teaching scenario, we should not require students to spend their attention on these subtleties, e.g. remembering the Coq theorem names of these different versions.

(b) Propositions about the membership relation are less intuitive to read.
Students are more willing to use standard math notations like $a \in X$ and $(a,b) \in X$ instead of \coqinline{X a} and \coqinline{X a b} respectively.
One potential solution is to define such membership relations using higher order functions:
\begin{lstlisting}[mathescape=true]
    Definition In1 (a: A) (X: A -> Prop) := X a.
    Notation "x $\in$ y" := (In1 x y).
    Definition In2 (p: A * B) (X: A -> B -> Prop) :=
      X (fst p) (snd p).
\end{lstlisting}
Then, we need a unified definition of \coqinline{In1}, \coqinline{In2} and potentially some \coqinline{In3} for ternary relations.

\paragraph{Remark.} Giving up curried Coq types like \coqinline{A -> B -> Prop} and choosing to use uncurried Coq types like \coqinline{A * B -> Prop} seems a potential solution to these two problems above.
However, this does not really work because we would be unable to use all standard library supports for relations, including those useful type classes (like \coqinline{Equivalence}) and proof automation (like the morphism-based \coqinline{rewrite} tactic).

\paragraph{Problems in set-related proof automation.}
When students learn set theory in their discrete math course, set equivalence and set inclusion can be reduced to propositions about the membership relation, i.e.
$$ X = Y  \  \ \ \text{iff.} \ \ \ \forall a. \ a \in X \iff a \in Y $$
$$ X \subseteq Y  \  \ \ \text{iff.} \ \ \ \forall a. \ a \in X \Rightarrow a \in Y $$
Such reduction is very useful because we can apply some proof automation like \coqinline{tauto} and \coqinline{firstorder} afterwards.
If we define set equivalence and set inclusion accordingly (see below), it seems that these reduction can be simply implemented as an \coqinline{unfold} tactic in Coq.
\begin{lstlisting}[mathescape=true]
    Definition included2 (X Y: A -> B -> Prop): Prop :=
      forall a b, X a b -> Y a b.
    Notation "x $\subseteq$ y" := (included2 x y).
\end{lstlisting}
But in fact, unfolding does not work, because it cannot reduces \coqinline{X $\subseteq$ Y} to
\begin{lstlisting}[mathescape=true]
    forall a b, (a, b) $\in$ X -> (a, b) $\in$ Y.
\end{lstlisting}
As we just mentioned, we want to use standard math symbols which is more intuitive to read.

One may argue that \coqinline{included2} should be defined using the membership relation \coqinline{In2}. But this cannot solve our problem either, since only changing \coqinline{included2}'s definition does not allow us to unfold \coqinline{X $\subseteq$ Y $\cup$ Z} to
\begin{lstlisting}[mathescape=true]
    forall a b, (a, b) $\in$ X -> (a, b) $\in$ Y \/ (a, b) $\in$ Z.
\end{lstlisting}

\paragraph{Problems about composing binary relations.}

If we only consider binary relation composition, it can be easily defined by FOL propositions (see below).
But we find that the composition operator between a binary relation and a unary relation (see below) shares many similar properties with the ordinary composition, e.g. associativity and distribution over set union.
\begin{lstlisting}[mathescape=true]
    Definition comp22 (R: A -> B -> Prop) (S: B -> C -> Prop) :=
      fun a c => exists b, R a b /\ S b c.
    Definition comp21 (R: A -> B -> Prop) (S: B -> Prop) :=
      fun a => exists b, R a b /\ S b.
\end{lstlisting}
This similarity between \coqinline{comp22} and \coqinline{comp21} is natural, because a unary relation (i.e. \coqinline{S: B -> Prop}) can be treated as a binary relation to the unit type (i.e. \coqinline{S: B -> unit -> Prop}).

We want to have a unified definition of these two composition operators, and even some other composition operator about labeled relations (see below), so that our formalized material can help students understand this similarity.
\begin{lstlisting}[mathescape=true]
    Definition comp33
      (R: A -> list event -> B -> Prop)
      (S: B -> list event -> C -> Prop) :=
    fun a l c => exists l1 l2 b,
      R a l1 b /\ S b l2 c /\ l = l1 ++ l2.
\end{lstlisting}

\paragraph{Our set library.}
In this paper, we present our Coq library of sets, which is initially designed for the purpose of teaching a programming language course.
The source files can be found at: \url{https://bitbucket.org/qinxiang-SJTU/sets/src/}.
During our development, we found that some of our requirement mentioned above like a unified definition of unions and a unified definition of relation compositions is also meaningful for professional users.
Thus we carefully design our unified definitions and automated tactics so that using our library will cause no proof-term overhead comparing with using naive definitions like \coqinline{union1}, \coqinline{union2}, \coqinline{comp22}, etc.

\paragraph{The rest of this paper.}
We will first give an overview of this library from the users' point of view in \S\ref{sec:overview}.
We will then introduce its application in teaching in \S\ref{sec:application}.
For techniques used in developing this library,
we introduce our unified definitions of set operators, relation compositions, and the membership relation in \S\ref{sec:def}.
We introduce our proof automation support in \S\ref{sect:unfold}.
Finally, we discuss related work in \S\ref{sect:related} and conclude in \S\ref{sect:concl}.


\section{Overview of the library} \label{sec:overview}

\subsection{Functions, predicates and constants}

In this library, every Coq type \coqinline{T} with form \coqinline{A -> B -> ... -> Prop} is treated as sets, and we provide the following functions, predicates and constants for \coqinline{T}.

\begin{lstlisting}[mathescape=true]
  Sets.equiv: T -> T -> Prop                              (* == *)
  Sets.included: T -> T -> Prop                           (* $\subseteq$ *)
  Sets.empty: T                                           (* $\emptyset$ *)
  Sets.full: T
  Sets.union: T -> T -> T                                 (* $\cup$ *)
  Sets.intersect: T -> T -> T                             (* $\cap$ *)
  Sets.indexed_union: forall {I: Type}, (I -> T) -> T     (* $\bigcup$ *)
  Sets.indexed_intersect: forall {I: Type}, (I -> T) -> T (* $\bigcap$ *)
  Sets.general_union: (T -> Prop) -> T
  Sets.general_intersect: (T -> Prop) -> T
\end{lstlisting}

Their meanings are mostly ordinary. \coqinline{Sets.indexed\_union $X$} and \coqinline{Sets.general\_union $P$} mean
$$\bigcup_{i\in I} X(i) \ \ \ \text{and} \ \ \ \bigcup_{X \ \text{satisfies} \ P}X$$
respectively. Similarly, \coqinline{Sets.indexed\_intersect $X$} and \coqinline{Sets.general\_intersect $P$} mean
$$\bigcap_{i\in I} X(i) \ \ \ \text{and} \ \ \ \bigcap_{X \ \text{satisfies} \ P}X$$
respectively. Thus, an instance of intersection's distribution law over a countably infinite union can be written in Coq as:
\begin{lstlisting}[mathescape=true]
  forall (A: Type) (X: A -> Prop) (Y: nat -> A -> Prop),
    X $\cap$ $\bigcup$ Y == $\bigcup$ (fun n => X $\cap$ Y n)
\end{lstlisting}

Besides, \coqinline{Sets.In} is an uncurried Coq function representing the membership relation, i.e. from users' point of view, \coqinline{Sets.In} ``have'' the following Coq types.
\begin{lstlisting}[mathescape=true]
  A -> (A -> Prop) -> Prop                       (* a $\in$ X *)
  A * B -> (A -> B -> Prop) -> Prop              (* (a, b) $\in$ X *)
  A * B * C -> (A -> B -> C -> Prop) -> Prop     (* (a, b, c) $\in$ X *)
  ...
\end{lstlisting}

The set library defines \coqinline{Rels.concat} (\coqinline{R $\circ$ S}) to represent different kinds of relation compositions. From users' point of view, the following variants are supported.
\begin{itemize}
\item If $R \subseteq A \times B$ and $S \subseteq B \times C$, then $R \circ S \subseteq A \times C$ (in Coq, \coqinline{R: A -> B -> Prop}, \coqinline{S: B -> C -> Prop}, and \coqinline{R $\circ$ S: A -> C -> Prop}) and
\begin{lstlisting}[mathescape=true]
  (a, c) $\in$ R $\circ$ S <->
    exists b: B, (a, b) $\in$ R /\ (b, c) $\in$ S.
\end{lstlisting}

\item If $R \subseteq A \times B$ and $S \subseteq B$, then $R \circ S \subseteq A$ (in Coq, \coqinline{R: A -> B -> Prop}, \coqinline{S: B -> Prop}, and \coqinline{R $\circ$ S: A -> Prop}) and
\begin{lstlisting}[mathescape=true]
  a $\in$ R $\circ$ S <->
    exists b: B, (a, b) $\in$ R /\ b $\in$ S.
\end{lstlisting}

\item If $R \subseteq A \times E^{*} \times B$ and $S \subseteq B \times E^{*} \times C$, then $R \circ S \subseteq A \times E^{*} \times C$ (in Coq, \coqinline{R: A -> list E -> B -> Prop}, \coqinline{S: B -> list E -> C -> Prop}, and \coqinline{R $\circ$ S: A -> list E -> C -> Prop}) and
\begin{lstlisting}[mathescape=true]
  (a, l, c) $\in$ R $\circ$ S <->
    exists (b: B) (l1 l2: list E),
      (a, l1, b) $\in$ R /\ (b, l2, c) $\in$ S /\ l = l1 ++ l2.
\end{lstlisting}

\item If $R \subseteq A \times E^{*} \times B$ and $S \subseteq B \times E^{*}$, then $R \circ S \subseteq A \times E^{*}$ (in Coq, \coqinline{R: A -> list E -> B -> Prop}, \coqinline{S: B -> list E -> Prop}, and \coqinline{R $\circ$ S: A -> list E -> Prop}) and
\begin{lstlisting}[mathescape=true]
  (a, l) $\in$ R $\circ$ S <->
    exists (b: B) (l1 l2: list E),
      (a, b) $\in$ R /\ b $\in$ S /\ l = l1 ++ l2.
\end{lstlisting}

\item If $R \subseteq A \times E^{*} \times B$ and $S \subseteq B \times E^{\omega}$, then $R \circ S \subseteq A \times E^{\omega}$ (in Coq, \coqinline{R: A -> list E -> B -> Prop}, \coqinline{S: B -> Stream E -> Prop}, and \coqinline{R $\circ$ S: A -> Stream E -> Prop}) and
\begin{lstlisting}[mathescape=true]
  (a, l) $\in$ R $\circ$ S <->
    exists (b: B) (l1: list E) (l2: Stream E),
      (a, b) $\in$ R /\ b $\in$ S /\ l = l1 +++ l2
\end{lstlisting}
where for any \coqinline{l: Stream E},
\begin{lstlisting}[mathescape=true]
  [e1; e2; ...; en] +++ l =
    Cons e1 (Cons e2 (... (Cons en l) ...)
\end{lstlisting}
\end{itemize}
Although we do provide supports for more complicated cases, they are not used in teaching.

The set library also defines \coqinline{Rels.id} to represent ``the identity relation''. The most useful variants are:
\begin{itemize}
\item For \coqinline{a b: A},
\begin{lstlisting}[mathescape=true]
  (a, b) $\in$ Rels.id <-> a = b.
\end{lstlisting}
\item For \coqinline{a b: A} and \coqinline{l: list E},
\begin{lstlisting}[mathescape=true]
  (a, l, b) $\in$ Rels.id <-> a = b /\ l = nil.
\end{lstlisting}
\end{itemize}

\subsection{Useful theorems}

We provide 61 useful lemmas about these set-related functions and predicates\footnote{Users usually cannot remember all of their names and thus need to seek help from Coq's \coqinline{Search} command.}.
Here are the most important ones.

\begin{itemize}
\item Commutativity and associativity of unions and intersections.
\begin{lstlisting}[mathescape=true]
  Sets_union_comm:
    forall (x y: T), x $\cup$ y == y $\cup$ x;
  Sets_union_assoc:
    forall (x y z: T), (x $\cup$ y) $\cup$ z == x $\cup$ (y $\cup$ z);
  Sets_intersect_comm:
    forall (x y: T), x $\cap$ y == y $\cap$ x;
  Sets_intersect_assoc:
    forall (x y z: T), (x $\cap$ y) $\cap$ z == x $\cap$ (y $\cap$ z);
\end{lstlisting}
where \coqinline{T} is a Coq type representing sets.

\item Associativity of relation composition.
\begin{lstlisting}[mathescape=true]
  Rels_concat_assoc:
    forall x y z, (x $\circ$ y) $\circ$ z == x $\circ$ (y $\circ$ z).
\end{lstlisting}
We support different variants of this theorem based on different variants of relation compositions. Specifically, \coqinline{x},  \coqinline{y} and \coqinline{z} above may have the following Coq types.
\begin{lstlisting}[mathescape=true]
  Case 1:
    x: A -> B -> Prop, y: B -> C -> Prop, z: C -> D -> Prop
  Case 2:
    x: A -> B -> Prop, y: B -> C -> Prop, z: C -> Prop
  Case 3:
    x: A -> list E -> B -> Prop, y: B -> list E -> C -> Prop,
    z: C -> list E -> D -> Prop
  Case 4:
    x: A -> list E -> B -> Prop, y: B -> list E -> C -> Prop,
    z: C -> list E -> Prop
  Case 5:
    x: A -> list E -> B -> Prop, y: B -> list E -> C -> Prop,
    z: C -> Stream E -> Prop
\end{lstlisting}

\item The distribution law of relation composition over unions.
\begin{lstlisting}[mathescape=true]
  Rels_concat_union_distr_r:
    forall x1 x2 y,
      (x1 $\cup$ x2) $\circ$ y == x1 $\circ$ y $\cup$ x2 $\circ$ y;
  Rels_concat_union_distr_l:
    forall x y1 y2,
      x $\circ$ (y1 $\cup$ y2) == x $\circ$ y1 $\cup$ x $\circ$ y2;
  Rels_concat_indexed_union_distr_r:
    forall xs y,
      $\bigcup$ xs $\circ$ y == $\bigcup$ (fun i => (xs i) $\circ$ y);
  Rels_concat_indexed_union_distr_l:
    forall x ys,
      x $\circ$ $\bigcup$ ys == $\bigcup$ (fun i => x $\circ$ (ys i)).
\end{lstlisting}
Similar to the associativity of relation composition, all different variants of relation composition have these distribution laws.
\end{itemize}

\subsection{Tactic support}

We provide ``\coqinline{Sets\_unfold}'' and ``\coqinline{Sets\_unfold in ...}'' for users to unfold set related definitions into a logical proposition about the membership relation.
The lower-case version \coqinline{sets\_unfold} has a similar functionality, but treats sets as ordinary curried Coq functions and does not generate ``$\in$'' in propositions.
For example, \coqinline{X $\subseteq$ Y $\cup$ Z} is transformed to the following two propositions by \coqinline{Sets\_unfold} and \coqinline{sets\_unfold} respectively.
\begin{lstlisting}[mathescape=true]
    forall a b, (a, b) $\in$ X -> (a, b) $\in$ Y \/ (a, b) $\in$ Z
    forall a b, X a b -> Y a b \/ Z a b
\end{lstlisting}
Mainly, \coqinline{Sets\_unfold} is designed for teaching scenarios and \coqinline{sets\_unfold} is designed for potential professional users.
These two are both zero-cost in the sense of Coq proof term size.

We prove that \coqinline{Sets.union}, \coqinline{Sets.intersect} and \coqinline{Rels.concat} preserve set equivalence and set inclusion.
These statements are formalized using the morphism type classes in Coq's standard libraries.
Thus, users can directly use Coq's built-in \coqinline{rewrite} tactic to handle set equivalence and set inclusion in proofs.

Application examples of these tactics can be found in the next section.

\section{Applications in teaching}\label{sec:application}

This set library is designed and used in teaching three different undergrad courses (for three distinct classes of students) about programming language theory in Shanghai Jiao Tong University, one of the top universities in China.

\begin{table}[htb]   
\begin{center}   
\begin{tabular}{|c|c|c|c|c|}   
\hline   \textbf{Course} & \textbf{No. of} & \textbf{Year and} & \textbf{Knowledge about} & \textbf{Programming} \\   
 \textbf{ID} & \textbf{students} & \textbf{semester} & \textbf{set theory} & \textbf{experiences} \\   
\hline   CS2612 & about 40 & 3rd year, fall & Yes &  2 years CS major courses \\ 
\hline   CS2205 & about 25 & 2nd year, fall & No &  1 year CS major courses \\ 
\hline   CS2206 & about 10 & 2nd year, spring & No &  more than 4 years \\ 
\hline   
\end{tabular}   
\end{center}   
\end{table}

Student backgrounds are slightly different in these three courses. 
Students in CS2612 have learnt set theory seriously.
Thus, they understand that a binary relation $R$ from $A$ to $B$ is actually a subset of the Cartesian product $A \times B$, and notations $aRb$ or $(a,b) \in R$ can be used to talk about $R$'s elements.
They have also learnt that binary relation composition has associativity and the distribution law over unions.
Students in CS2205 and CS2206 have not learnt anything about binary relations but they have learnt the semantic theory of first order logic in a logic course.
Thus, they have a brief understanding of relations.
For all of these three courses, students have no prior knowledge about the Coq proof assistant.

The purpose of using Coq in these three courses are not making students Coq experts but helping students understand definitions and proofs.

\paragraph{For understanding binary relations.}

\begin{figure}[htbp]
      \centering
      \includegraphics[scale=0.3]{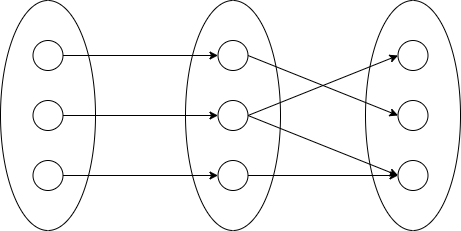}
      \caption{Diagrams for binary relation composition}
      \label{fig:comp}
\end{figure}

In CS2205 and CS2206, the instructor taught basic concepts about binary relations and the composition operation on the blackboard.
Some diagrams like figure \ref{fig:comp} were used for describing the intuition behind.
Properties like composition's associativity and distribution law over unions were first introduced with the help of such diagrams,
then formally proved using Coq.
The instructor would type a Coq proof for one theorem like the following in class. This proof would only use binary relation composition's definition just taught on the blackboard, and all implementation of our set library is hidden.
\begin{lstlisting}[mathescape=true]
Theorem Rels_concat_assoc:
  forall
    (A B C D: Type)
    (R1: A -> B -> Prop) (R2: B -> C -> Prop) (R3: C -> D -> Prop),
  (R1 $\circ$ R2) $\circ$ R3 == R1 $\circ$ (R2 $\circ$ R3).
Proof.
  intros. Sets_unfold. intros a d.
  (** Now, the proof goal is:
       (exists c : C,
         (exists b : B, (a, b) $\in$ R1 /\ (b, c) $\in$ R2) /\
         (c, d) $\in$ R3) <->
       (exists b : B,
         (a, b) $\in$ R1 /\ 
         (exists c : C, (b, c) $\in$ R2 /\ (c, d) $\in$ R3)) *)
  (** The rest is a FOL proof. *)
  split.
  + intros [c [[b [? ?]] ?]].
    exists b.
    split.
    - tauto.
    - exists c. tauto.
  + intros [b [? [c [? ?]]]].
    exists c.
    split.
    - exists b. tauto.
    - tauto.
Qed.
\end{lstlisting}
Students were then asked to formally prove another theorem by themselves.
We find that Coq helps students understand how to really prove something using math definitions.

In CS2612, we use Coq to help students review basic concepts about binary relations that students had already learnt.
The class would skim those important properties very quickly and the instructor would re-explain the intuition on the blackboard.
Then, Coq was used to demonstrate a formal proof of binary relation composition's associativity so that students could establish a connection between intuitive understanding and strict math proofs during the review.

\paragraph{For defining denotational semantics.}

In a typical textbook denotational semantics, a program command's denotation is a binary relation from initial program states to ending states, i.e. $\denote{c} \subseteq \mathrm{state} \times \mathrm{state}$.
Then,
\begin{eqnarray*}
\denote{c_1; c_2} &=& \denote{c_1} \circ \denote{c_2} \\
\denote{\ifthenelse{e}{c_1}{c_2}} &=& 
\mathrm{test\_true}(\denote{e}) \circ \denote{c_1} \ \cup \ \mathrm{test\_false}(\denote{e}) \circ \denote{c_2}
\end{eqnarray*}
where expressions are assumed to be side-condition-free and
\begin{eqnarray*}
(s_1, s_2) \in \mathrm{test\_true}(\denote{e}) &\text{iff.}&
s_1 = s_2 \ \text{and $e$ is true on $s_1$;} \\
(s_1, s_2) \in \mathrm{test\_false}(\denote{e}) &\text{iff.}&
s_1 = s_2 \ \text{and $e$ is false on $s_1$.}
\end{eqnarray*}

Then, $(c_1; c_2); c_3$ and $c_1; (c_2; c_3)$ are semantically equivalent because
$$\denote{(c_1; c_2); c_3} = 
(\denote{c_1} \circ \denote{c_2}) \circ \denote{c_3} = 
\denote{c_1} \circ (\denote{c_2} \circ \denote{c_3}) =
\denote{c_1; (c_2; c_3)}.$$
Also, $(\text{if} \ (e) \ \text{then} \ c_1 \ \text{else} \ c_2); c_3$ and $\text{if} \ (e) \ \text{then} \ (c_1; c_3) \ \text{else} \ (c_2; c_3)$ are semantically equivalent because
\begin{eqnarray*}
 && \denote{(\text{if} \ (e) \ \text{then} \ c_1 \ \text{else} \ c_2); c_3} \\
 &=& (\mathrm{test\_true}(\denote{e}) \circ \denote{c_1} \ \cup \ \mathrm{test\_false}(\denote{e}) \circ \denote{c_2}) \circ \denote{c_3} \\
 &=& \mathrm{test\_true}(\denote{e}) \circ \denote{c_1} \circ \denote{c_3}\ \cup \ \mathrm{test\_false}(\denote{e}) \circ \denote{c_2} \circ \denote{c_3} \\  
 &=& \denote{\text{if} \ (e) \ \text{then} \ (c_1; c_3) \ \text{else} \ (c_2; c_3)}
\end{eqnarray*}
In the third line of this proof above, the associativity of binary relation composition is implicitly applied!
A Coq proof used in class (see below) reminds everyone this fact.
\begin{lstlisting}[mathescape=true]
Lemma if_seq: forall e c1 c2 c3,
  $\llbracket$ (if e then c1 else c2); c3 $\rrbracket$ ==
  $\llbracket$ if e then (c1; c3) else (c2; c3) $\rrbracket$.
Proof.
  intros. simpl; unfold seq_sem, if_sem.
  (** Now, the proof goal is:
        (test_true $\llbracket$ e $\rrbracket$ $\circ$ $\llbracket$ c1 $\rrbracket$ $\cup$ test_false $\llbracket$ e $\rrbracket$ $\circ$ $\llbracket$ c2 $\rrbracket$) $\circ$ $\llbracket$ c3 $\rrbracket$
        == test_true $\llbracket$ e $\rrbracket$ $\circ$ ($\llbracket$ c1 $\rrbracket$  $\circ$ $\llbracket$ c3 $\rrbracket$) $\cup$
           test_false $\llbracket$ e $\rrbracket$ $\circ$ ($\llbracket$ c1 $\rrbracket$ $\circ$ $\llbracket$ c3 $\rrbracket$) *)
  rewrite Rels_concat_union_distr_r.
  rewrite !Rels_concat_assoc.
  (** This line above uses the associativity. *)
  reflexivity.
Qed.
\end{lstlisting}

\paragraph{For denotational semantics of more practical languages.}

Using Coq also allows us to discuss more practical language features in these programming language courses.
Specifically, when we present a realistic semantics with many subtleties involved, we can check line by line whether the original concise theory is broken or not.
In these course,
we provide extended and optional reading material about nondeterministic programming languages and program languages with observable behaviors (like IO events) for interested students.

For nondeterminism, we follow Back's approach \cite{Back83}, i.e.
\begin{itemize}
\item $\denote{c}.(\mathrm{nrm})$ is a binary relations between program states; $(s_1, s_2) \in \denote{c}.(\mathrm{nrm})$ iff. executing $c$ from the initial state $s_1$ may terminate on $s_2$;
\item $\denote{c}.(\mathrm{inf})$ is a subset of program states; $s \in \denote{c}.(\mathrm{inf})$ iff. executing $c$ from the initial state $s$ may not terminate.
\end{itemize}
Then,
\begin{eqnarray*}
\denote{c_1; c_2}.(\mathrm{inf}) &=& \denote{c_1}.(\mathrm{inf}) \ \ \cup \ \ \denote{c_1}.(\mathrm{nrm}) \circ \denote{c_2}.(\mathrm{inf}) \\
\denote{\ifthenelse{e}{c_1}{c_2}}.(\mathrm{inf}) &=& 
\mathrm{test\_true}(\denote{e}) \circ \denote{c_1}.(\mathrm{inf}) \ \ \cup \ \ \mathrm{test\_false}(\denote{e}) \circ \denote{c_2}.(\mathrm{inf})
\end{eqnarray*}
Our set library reloads the ``$\circ$'' operator, which makes this definition above natural and intuitive.

For programs with observable behaviors, $\denote{c} \subseteq \mathrm{state} \times \mathrm{list\_of\_events} \times \mathrm{state}$.
Specifically, $(s_1, \tau, s_2) \in \denote{c}$ if and only if executing $c$ from $s_1$ will generate event trace $\tau$ and terminate in state $s_2$.
Again, we have $\denote{c_1; c_2} = \denote{c_1} \circ \denote{c_2}$ according to the meaning of our reloaded ``$\circ$''.

\paragraph{Applying fixed point theorems.}

The fixed point theorem part is another place where students significantly benefit from using Coq.
For example, in order to prove $f(x) = x$ on a partial order, it suffices to show that $f(x) \le x$ and $x \le f(x)$.
But usually, students will not ask themselves why.
It easily happen that they forgot they are working on a partial order not on real number's ordering.
Coq forces students to answer such questions in every step in their homework proofs. For example, the answer to the question above is anti-symmetry.

The set library is used when proving $(P(\mathrm{state} \times \mathrm{state}), \subseteq)$ is a complete lattice, i.e. it is a partial order and every set of binary relations has a least upper bound (in a set theory sense).
The instructor would check some of these properties in class, and others are left as homework --- this is a traditional design in a programming language theory course but now it is done in Coq.
For example, the following is the Coq proof of $\subseteq$'s anti-symmetry.

\begin{lstlisting}[mathescape=true]
Theorem relation_inclusion_antisymm:
  forall (x y: state -> state -> Prop),
    x $\subseteq$ y -> y $\subseteq$ x -> x == y.
Proof.
  Sets_unfold. intros.
  (** Now, the proof goal is
        x, y : state -> state -> Prop
        H : forall a a0 : state, (a, a0) $\in$ x -> (a, a0) $\in$ y
        H0 : forall a a0 : state, (a, a0) $\in$ y -> (a, a0) $\in$ x
        a, a0 : state
        ------------------------------------
        (a, a0) $\in$ x <-> (a, a0) $\in$ y *)
  specialize (H a a0).
  specialize (H0 a a0).
  tauto.
Qed.
\end{lstlisting}

\paragraph{Summary and evaluation.}
Overall, using the Coq proof assistant helps us clarify potential ambiguity in natural language description, and emphasize important proof steps that may be ignored.
Also, it allows us to extend typical toy languages to more realistic settings in theory courses --- using the proof assistant gives students confidence that we do cover all subtle cases in proofs.

On the negative side, however, using Coq necessitate dedicating a few lectures to instructing students how to use the proof assistant itself.
We try to minimize it in our course design.
We teach students Coq inductive type when introducing programming language AST.
We teach students structural recursive function and structure inductive proof when teaching denotational semantics and its theory.
We teach students induction proofs over natural numbers in Coq when teaching Kleen fixed point theorem\footnote{We teach students general Coq inductive type quite early but the fact that natural numbers are defined as a specific inductive type in Coq are postponed to the first time that we need an induction proof over natural numbers.}.
Coq proofs about logic (conjunction, disjunction, implication, negation and quantifiers) are taught separately.
We choose not to teach students Curry-Howard correspondence, and not to teach students Coq's inductively defined propositions --- we find that they are unnecessary for the courses and hard for students to understand.

Designing the set library is one of the effort for compressing the time spending on Coq itself.
We use this library to hide how set-related operators are implemented and encoded as Coq definitions.
Students find that this library is very easy to use and related Coq proofs that we demonstrated in classes are very easy to understand.

\section{Implementation of the library: unified definitions}\label{sec:def}

\subsection{Set operators}
\label{sect:op}

We define union, intersection, empty set, full set, general union, general intersection, set equivalence and set inclusion using Coq's type class system.
The main idea is, if \coqinline{T} is a Coq type of form \coqinline{A -> B -> ... -> Prop}, then we can construct a instance of type class \coqinline{Sets.SETS T}.

\begin{lstlisting}[mathescape=true]
    Module Sets.
      Class SETS (T: Type): Type := {
        equiv: T -> T -> Prop;
        included: T -> T -> Prop;
        empty: T;
        full: T;
        union: T -> T -> T;
        indexed_union: forall {I: Type}, (I -> T) -> T;
        ...
      }.
    End Sets.
\end{lstlisting}
Constructing instances of \coqinline{Sets.SETS} only needs the following conclusions.
\begin{lstlisting}[mathescape=true]
    Instance Prop_SETS: Sets.SETS Prop.
    Instance lift_SETS A B: Sets.SETS B -> Sets.SETS (A -> B).
\end{lstlisting}
We put some primary properties of these set operators in another type class beside \coqinline{Sets.SETS} and derived other properties from primary properties.
Overall, defining these operators and proving these properties is not hard.
Researchers have achieved that when formalizing shallowly embedded assertion languages.

\subsection{Compositions}
\label{sect:comp}

We formalize general composition using Coq type class \coqinline{Rels.RELS} and an auxiliary type class \coqinline{Rels.RELS}.
\begin{lstlisting}[mathescape=true]
    Module Rels.
      Class PRE_RELS (A T1 T2 T: Type): Type :=
      { concat_aux: T1 -> T2 -> A -> T }.
      Class RELS (T1 T2 T: Type): Type :=
      { concat: T1 -> T2 -> T }.
    End Rels.
\end{lstlisting}
For example, the ordinary composition \coqinline{comp22} can be defined by an instance \coqinline{R2} of
\begin{lstlisting}[mathescape=true]
    Rels.RELS (A -> B -> Prop) (B -> C -> Prop) (A -> C -> Prop)
\end{lstlisting}
which is constructed by an instance \coqinline{R1} of
\begin{lstlisting}[mathescape=true]
    Rels.PRE_RELS B (B -> Prop) (C -> Prop) (C -> Prop).
\end{lstlisting}
To be more specific,
\begin{lstlisting}[mathescape=true]
    @concat_aux _ _ _ _ R1 :=
      fun (X: B -> Prop) (Y: C -> Prop) =>
        fun (b: B) (c: C) => X b /\ Y c.
    @concat _ _ _ R2 :=
      fun (X: A -> B -> Prop) (Y: B -> C -> Prop)
        fun (a: A) =>
          Sets.indexed_union (fun b => concat_aux (X a) (Y b) b)
     (* which equals to
      fun (X: A -> B -> Prop) (Y: B -> C -> Prop)
        fun (a: A) (c: C) =>
          exists b: B, concat_aux (X a) (Y b) b *)
\end{lstlisting}
Similarly, \coqinline{comp21} can be defined by an instance \coqinline{R2'} of
\begin{lstlisting}[mathescape=true]
    Rels.RELS (A -> B -> Prop) (B -> Prop) (A -> Prop)
\end{lstlisting}
which is constructed by an instance \coqinline{R1'} of
\begin{lstlisting}[mathescape=true]
    Rels.PRE_RELS B (B -> Prop) Prop Prop.
\end{lstlisting}
We prove related properies based on this type class system.

\subsection{The membership relation}
\label{sect:in}

We define our general membership relation by two type classes.

\begin{lstlisting}[mathescape=true]
    Module SetsEle.
      Class PRE_SETS_ELE (T RES E: Type): Type := ...
      Class SETS_ELE (T E: Type): Type :=
      { In: E -> T -> Prop;
        set_transfer: T -> T; }.
    End SetsEle.
\end{lstlisting}
As a result, \coqinline{In1} and \coqinline{In2} mentioned in \S\ref{sect:introduction} are equivalent to
\begin{lstlisting}[mathescape=true]
    @SetsEle.In (A -> Prop) A SE1
    @SetsEle.In (A -> Prop) (A * B) SE2
\end{lstlisting}
respectively, where \coqinline{SE1} and \coqinline{SE2} are some instances.
We omit details here since the main techniques are the same as defining \coqinline{Rels.concat} --- \coqinline{SetsEle.In} in the type class \coqinline{SETS\_ELE} is the definition behind the notation of ``$\in$'' and instances of \coqinline{PRE\_SETS\_ELE} are auxiliaries used for building instances of \coqinline{SETS\_ELE}.
The function \coqinline{SetsEle.set\_transfer} has nothing to do with the implementation of \coqinline{SetsEle.In}.
It is used in the unfolding tactics (see \S\ref{sect:unfold}).

\section{Implementation of the library: the unfolding tactic}
\label{sect:unfold}

We have mentioned in \S\ref{sect:introduction} that simply unfolding Coq definitions of set equivalence and set inclusion cannot generate intuitive propositions.
For example, if \coqinline{X Y Z: A -> B -> Prop}, then unfolding \coqinline{X} $\subseteq$ \coqinline{Y} $\cup$ \coqinline{Z} results in
\begin{lstlisting}[mathescape=true]
    forall a b, X a b -> Y a b \/ Z a b.
\end{lstlisting}
Interestingly, unfolding is not too far way from our final solution! We only need to replace \coqinline{X}, \coqinline{Y} and \coqinline{Z} with Coq terms like
(\coqinline{fun a b => (a, b) $\in$ X}) first, then unfolding the definitions of $\subseteq$ and $\cup$ in
\begin{lstlisting}[mathescape=true]
    (fun a b => (a, b) $\in$ X) $\subseteq$
      (fun a b => (a, b) $\in$ Y) $\cup$ (fun a b => (a, b) $\in$ Z)
\end{lstlisting}
will generate the following proposition.
\begin{lstlisting}[mathescape=true]
    forall a b, (a, b) $\in$ X -> (a, b) $\in$ Y \/ (a, b) $\in$ Z.
\end{lstlisting}

The key in this transformation is turning Coq variables like \coqinline{X} into (\coqinline{fun a b => (a, b) $\in$ X}).
It is defined by \coqinline{SetsEle.set\_transfer} mentioned previously.
The effect of \coqinline{SetsEle.set\_transfer} is:
\begin{itemize}
\item If \coqinline{X: A -> Prop}, then
\coqinline{SetsEle.set\_transfer X} can be unfolded to
\begin{lstlisting}[mathescape=true]
    fun a => a $\in$ X
\end{lstlisting}

\item If \coqinline{X: A -> B -> Prop}, then
\coqinline{SetsEle.set\_transfer X} can be unfolded to
\begin{lstlisting}[mathescape=true]
    fun a b => (a, b) $\in$ X
\end{lstlisting}

\item If \coqinline{X: A -> B -> C -> Prop}, then
\coqinline{SetsEle.set\_transfer X} can be unfolded to
\begin{lstlisting}[mathescape=true]
    fun a b c => (a, b, c) $\in$ X
\end{lstlisting}

\item ...
\end{itemize}
It is worth mentioning, \coqinline{SetsEle.set\_transfer X} is $\beta\iota\delta$-equivalent to \coqinline{X} for every instance of \coqinline{SetsEle.} \coqinline{set\_transfer}.
That means, we can always
\begin{lstlisting}[mathescape=true]
    change X with (SetsEle.set_transfer X)
\end{lstlisting}
in Coq to complete the transformation as long as \coqinline{X} is a set.
We omit the implementation of \coqinline{SetsEle.set\_transfer} in this paper, and it uses the same technique used in the definition of \coqinline{Rels.concat}.

Our tactic \coqinline{Sets\_unfold} is an Ltac program in Coq.
It first marks all Coq variables which \coqinline{SetsEle.set\_transfer} should apply on, then uses a \coqinline{change} command to insert those \coqinline{SetsEle.set\_transfer}, and finally unfold the definition of \coqinline{SetsEle.set\_transfer} and other set-related definitions like unions and intersections.
\coqinline{Sets\_unfold} is very carefully implemented so that it causes zero proof-term overhead, i.e. it only replaces a Coq expression with a $\beta\iota\delta$-equivalent one, and its implementation cannot even include one single \coqinline{rewrite}.
For example, consider the following proof in Coq.
\begin{lstlisting}[mathescape=true]
  Example Sets_union_comm_included:
    forall (A: Type) (X Y: A -> Prop),
      X $\cup$ Y $\subseteq$ Y $\cup$ X.
  Proof.
    intros. Sets_unfold.
    intros. tauto.
  Qed.
\end{lstlisting}
Its proof term in Coq is:
\begin{lstlisting}[mathescape=true]
  fun (A : Type) (X Y : A -> Prop) (a : A) (H : a $\in$ X \/ a $\in$ Y) =>
    or_ind
      (fun H0 : a $\in$ X => or_intror H0)
      (fun H0 : a $\in$ Y => or_introl H0) H.
\end{lstlisting}
There is no redundancy compared to a manual proof.

\section{Related work}
\label{sect:related}

\paragraph{Formalization of sets.}
In Coq standard library, \coqinline{Ensemble X} is used to represent all subsets of \coqinline{X}, and is defined as \coqinline{X -> Prop} \cite{CoqSets}.
In Lean \cite{Lean} standard library, \coqinline{set X} is defined as \coqinline{X -> Prop} \cite{LeanSets}.
In Isabelle/HOL, \coqinline{set X} is isomorphic with \coqinline{X -> bool} \cite{IsabelleSets}.
These three definitions are very similar and none of them can provide a unified definition of unions and intersections for sets and relations as we proposed in our set library.
Besides, Coq standard library provides efficient implementations (based on lists and/or trees) of finite sets in the FSet module \cite{FSets} and the MSet module \cite{MSets}.
Since sets used in denotational semantics are not necessarily finite, FSet and MSet are not useful for us.

As mentioned in \S\ref{sect:op}, defining set intersection and set union is like defining conjunction and disjunction in shallowly embedded assertion languages.
Also, defining indexed set intersection and indexed set union  (\coqinline{Sets.indexed\_union} and \coqinline{Sets.indexed\_intersect} in our set library) is like defining existential quantifier and universal quantifier.
Specifically, an assertion over program states can be treated a subset of program states. If a program state is a pair of stack and heap, then assertions can be defined as \coqinline{stack -> heap -> Prop} in Coq.
The existing works that provide unified formalization of conjunction, disjunction and quantifiers for assertion languages include the MSL library in VST \cite{VSTBook2014} and the ModuRes library \cite{ModuRes} used in the Iris project \cite{IrisJournal}.
In comparison, our formalization of sets also provide definitions and proofs about relation compositions ``$\circ$'', and the membership relation ``$\in$''.

\paragraph{Formalization of denotational semantics.}
In many textbook about program semantics (like Winskel's \emph{Formal Semantics for Programming Languages: An Introduction} \cite{WinskelBook}), an imperative program's denotation is defined as a binary relation between program states.
Nipkow's formalization of denotational semantics follows this approach \cite{NipkowBook}.
A typical extension of such textbook denotational semantics is to additionally consider possible errors like null-pointer dereference, so a program's denotation becomes a binary relation between \coqinline{state} $\cup \ \{ \bot \}$, where $\bot$ represents errors.
Imperative-HOL \cite{ImperativeHOL} in Isabelle adopts this approach.

Our set library is used in teaching typical textbook denotations.
For potential program errors, our formalization is more close to Park's denotational semantics \cite{Park79}, dividing a program's denotation into several fields (like $\denote{c}.(\text{nrm})$, $\denote{c}.(\text{err})$ and $\denote{c}.(\text{err})$).
Although Park's original work is to find a theory for defining nondeterministic programs behavior including possible nontermination which fails main stream power domain theory before that \cite{DBLP:journals/jcss/Smyth78,DBLP:journals/siamcomp/Plotkin76}, we choose not to go that far in our introductory courses.

\paragraph{Comparison between denotational semantics and big step operational semantics.}
The \emph{Software Foundations} \cite{SF} book teaches big step operational semantics instead of denotational semantics.
Leroy's Mechanized Semantics course \cite{XavierCourse} also chooses to teach big step operational semantics instead of denotational semantics.
In comparison, denotational semantics and big step semantics are similar --- we say a program's denotation is a binary relation between program states, or a program evaluates to a binary relation between program states.
The main difference is: a program's denotation should be defined by a recursion over programs' syntax tree but a program's big step operational semantics is usually formalized by an inductive proposition, i.e. a rule-based definition in math.
In our courses, we choose not to include Coq inductive propositions.

\paragraph{Using proof assistant in teaching semantics.}
Interactive theorem provers are useful for teaching basic logical concepts.
Avigad shared his experience of teaching logics and set theories using Lean \cite{AvigadTeaching} in 2019.
His course material directly uses Lean's internal formalization of sets.
Students were taught to use Lean proving simple properties of sets.

\emph{Software Foundations} \cite{SF} is a famous textbook for teaching students to use the Coq theorem prover and for teaching basic programming language theories based on Coq.
Its material contains big step operational semantics, small step operational semantics and Hoare logic.
However, using \emph{Software Foundations} as a textbook means spending a lot of time teaching students to use Coq and then we have to cut down a decent portion of programming language theories from syllabus. 

Nipkow shared his teaching experience in VMCAI 2012 \cite{NipkowTeaching}.
He taught program semantics courses with the help of the Isabelle proof assistant.
His major points include
\begin{itemize}
\item \emph{``Teach Semantics, Not Proof Assistants.''}
\item \emph{``Teach Proofs, Not Proof Scripts.''}
\item \emph{``Do Not Let the Proof Assistant Dominate Your Presentation.''}
\end{itemize}
We adopt a similar philosophy in the preparation of our course material.

\section{Conclusion}
\label{sect:concl}

This paper describes a Coq library of sets and relations for teaching denotational semantics.
This library provides unified definitions of set operators, intuitive notation for the membership relation and automated tactics.
At the same time, we keep our definitions compatible with the existing formal definitions in Coq standard libraries.
With this set library, we can use the Coq proof assistant to help students understand denotational semantics, without spending too much time teaching how to use the theorem prover itself.
Also, some components of this library (like the unified definitions of set operators) are also useful for professional users of Coq when building denotational semantics.

\bibliographystyle{eptcs}
\bibliography{base}


\end{document}